\documentclass{PoS}

\title{J/$\psi$ polarization measurements in $p$+$p$ collisions at $\sqrt{s}$ = 200 and 500 GeV with the STAR experiment}

\ShortTitle{J/$\psi$ polarization measurements in $p$+$p$ collisions with the STAR experiment}

\author{\speaker{Barbara Trzeciak \thanks{For the STAR Collaboration} \thanks{Now at Faculty of Science, Utrecht University, Princetonplein 5, 3584 CC Utrecht, The Netherlands}}
        \\Faculty of Nuclear Sciences and Physical Engineering \\
		Czech Technical University in Prague \\
		Brehova 7, 115 19 Prague 1, Czech Republic \\
        E-mail: \email{b.a.trzeciak (at) uu.nl}}

\abstract{
In these proceedings, measurements of J/$\psi$ polarization in $p+p$ collisions at $\sqrt{s}$ = 200 and 500 GeV via the dielectron decay channel at mid-rapidity with the STAR experiment are discussed. At $\sqrt{s}$ = 200 GeV the polarization parameter, $\lambda_{\theta}$, related to the polar anisotropy is obtained in the helicity frame as a function of transverse momentum, 2 $< p_{T} <$ 6 GeV/$c$, and compared to different model predictions. 
A new J/$\psi$ polarization measurement at $\sqrt{s}$ = 500 GeV extends the previous analysis to a wide transverse momentum range of 5 $< p_{T} <$ 16 GeV/$c$. Also, the polarization parameter related to the azimuthal anisotropy, $\lambda_{\phi}$, is extracted in addition to $\lambda_{\theta}$, in two reference frames: the helicity and Collins-Soper frames. This allows for the frame invariant parameter calculation {\it{vs}} $p_{T}$ in these two frames.
}

\FullConference{The European Physical Society Conference on High Energy Physics\\
		22--29 July 2015\\
		Vienna, Austria}

\begin{document}
 
\section{Introduction}

J/$\psi$ is a bound state of charm ($c$) and anti-charm ($\overline{c}$) quarks. Final charmonium states have to be colorless, however they can be formed via a color-singlet (CS) or color-octet (CO) intermediate $c\overline{c}$ state. 
Based on this, there are different models that try to describe $c\overline{c}$ pair production and hadronization to the physical charmonium state, such as Color Singlet Model (CSM), Color Evaporation Model (CEM), or more sophisticated Non-Relativistic QCD (NRQCD) calculations.
CSM assumes that J/$\psi$ is created through the color-singlet intermediate state only, with the same quantum numbers as the final charmonium state. In the NRQCD approach $c\overline{c}$ color-octet intermediate states, in addition to the color-singlet states, can bind to form charmonia. 

The charmonium production mechanism in elementary particle collisions is still not yet exactly known.
For many years measurements of the J/$\psi$ cross section have been used to test different J/$\psi$ production models. While many models can describe relatively well the experimental data on the J/$\psi$ cross section in $p+p$ collisions \cite{Abelev:2009qaa, Adamczyk:2012ey, Adare:2011vq, PhysRevLett.79.572, Acosta:2004yw, Aad:2011sp, Khachatryan:2010yr, Aaij:2011jh}, they have different predictions for the J/$\psi$ polarization. Therefore, measurements of the J/$\psi$ polarization may allow to discriminate among different models and provide new insight into the J/$\psi$ production mechanism.

\section{J/$\psi$ polarization measurements with the STAR experiment}
\label{sec:measurements}

In STAR, the J/$\psi$ polarization has been analyzed at mid-rapidity at $\sqrt{s}$ = 200 and 500 GeV using the J/$\psi$ di-electron decay channel. The STAR detector~\cite{Ackermann:2002ad} is a multi-purpose detector that has a large acceptance at mid-rapidity, $\vert \eta \vert <$ 1, with a full azimuthal coverage. 
The Time Projection Chamber (TPC)~\cite{Anderson:2003ur} is the main tracking system and is used to identify particles via the ionization energy loss ($dE/dx$) measurement.
Identification of high-$p_{T}$ J/$\psi$ is possible using the Barrel Electromagnetic Calorimeter (BEMC) \cite{Beddo:2002zx} which allows to trigger on high-$p_{T}$ electrons and that improves high-$p_{T}$ electron identification.
Furthermore, the Time Of Flight (TOF) detector~\cite{Llope:2012zz} enhances the electron identification capability at low momenta where the $dE/dx$ bands for electrons and hadrons overlap. 

\subsection{Method}

J/$\psi$ polarization is measured via the angular distribution of the electrons from J/$\psi$ decay:
\begin{equation}
\frac{d^{2}N}{d(\cos\theta)d\phi} \propto 1+\lambda_\theta \cos^2\theta + \lambda_\phi \sin^2\theta \cos2\phi + \lambda_{\theta\phi}\sin2\theta \cos\phi
\end{equation}
\label{eq:decayD}
where $\theta$ and $\phi$ are polar and azimuthal angles, respectively; $\lambda_\theta$, $\lambda_\phi$ and $\lambda_{\theta\phi}$ are the angular decay coefficients.
The polar angle $\theta$ is defined as an angle between the positron momentum vector in the $J/\psi$ rest frame and a chosen polarization axis.
When integrated over $\phi$ or $\cos\theta$, the decay angular distribution has following forms:
\begin{equation}
W(\cos \theta) \propto 1 + \lambda _{\theta} \cos^{2}\theta
\label{eq:decayD_cos}
\end{equation}

\begin{equation}
W(\phi) \propto 1 + \frac{2\lambda_{\phi}}{3+\lambda_{\theta}} \cos2\phi	
\label{eq:decayD_phi}
\end{equation}
respectively.

In STAR inclusive J/$\psi$ polarization has been measured in two reference frames: the helicity (HX) and Collins-Soper (CS) frames. In the HX frame the polarization axis is defined along the J/$\psi$ momentum in the center-of-mass frame of colliding beams, and in the CS frame the polarization axis is a bisector of the angle formed by one beam direction and the opposite direction of the other beam, in the J/$\psi$ rest frame.
Furthermore, since values of measured polarization parameters depend on a chosen polarization axis, a frame invariant parameter has been proposed~\cite{Faccioli:2010kd}:
\begin{equation}
\lambda _{inv} = \frac{\lambda _{\theta} + 3 \lambda _{\phi}}{1-\lambda _{\phi}}
\end{equation}
Value of the parameter is independent of the chosen reference frame and so is a very good cross-check of measurements performed in different frames.

\section{J/$\psi$ polarization results}

\begin{figure}[ht]
		\centering
		\includegraphics[width=0.49\textwidth]{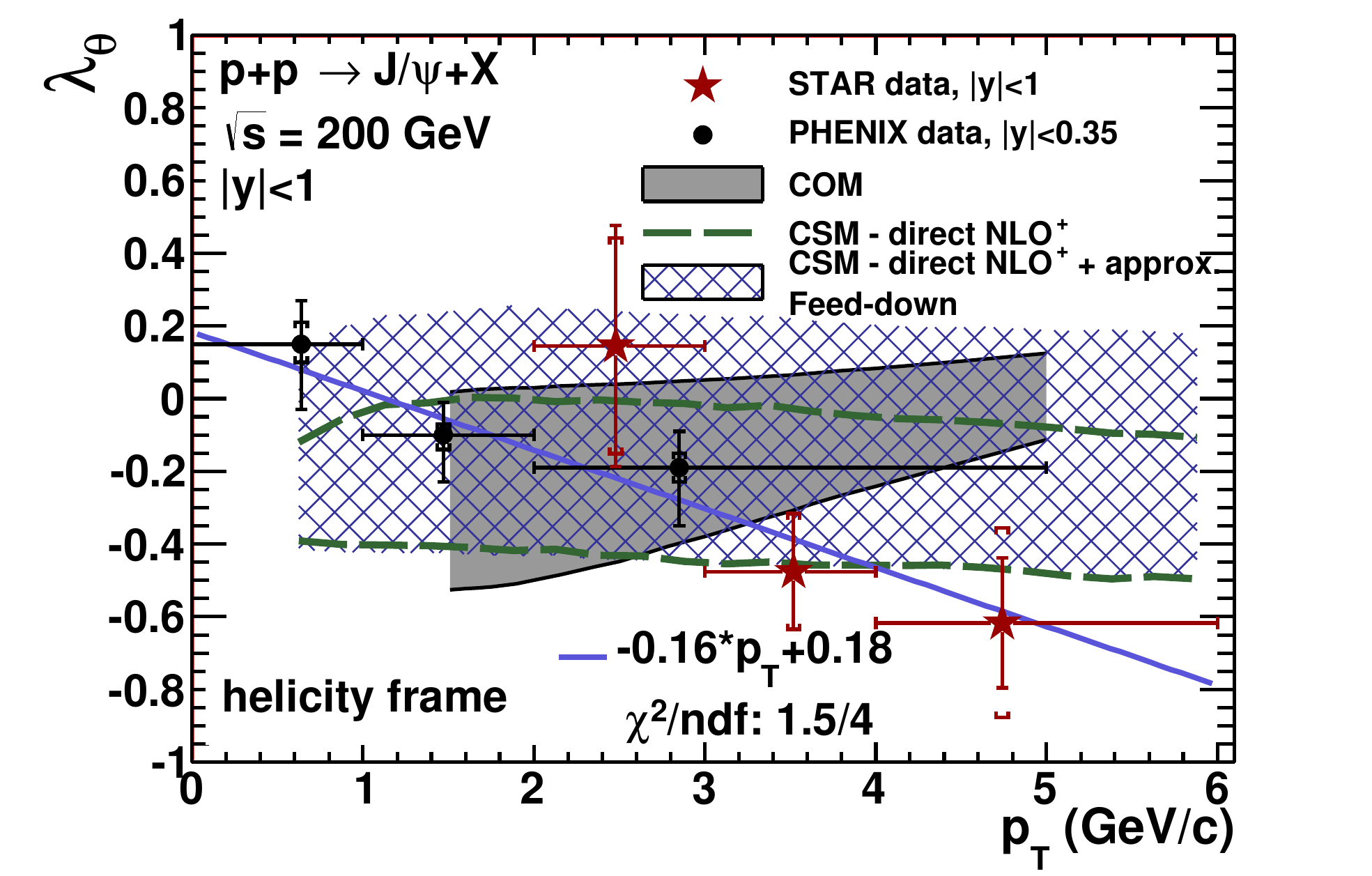}
		\includegraphics[width=0.49\linewidth]{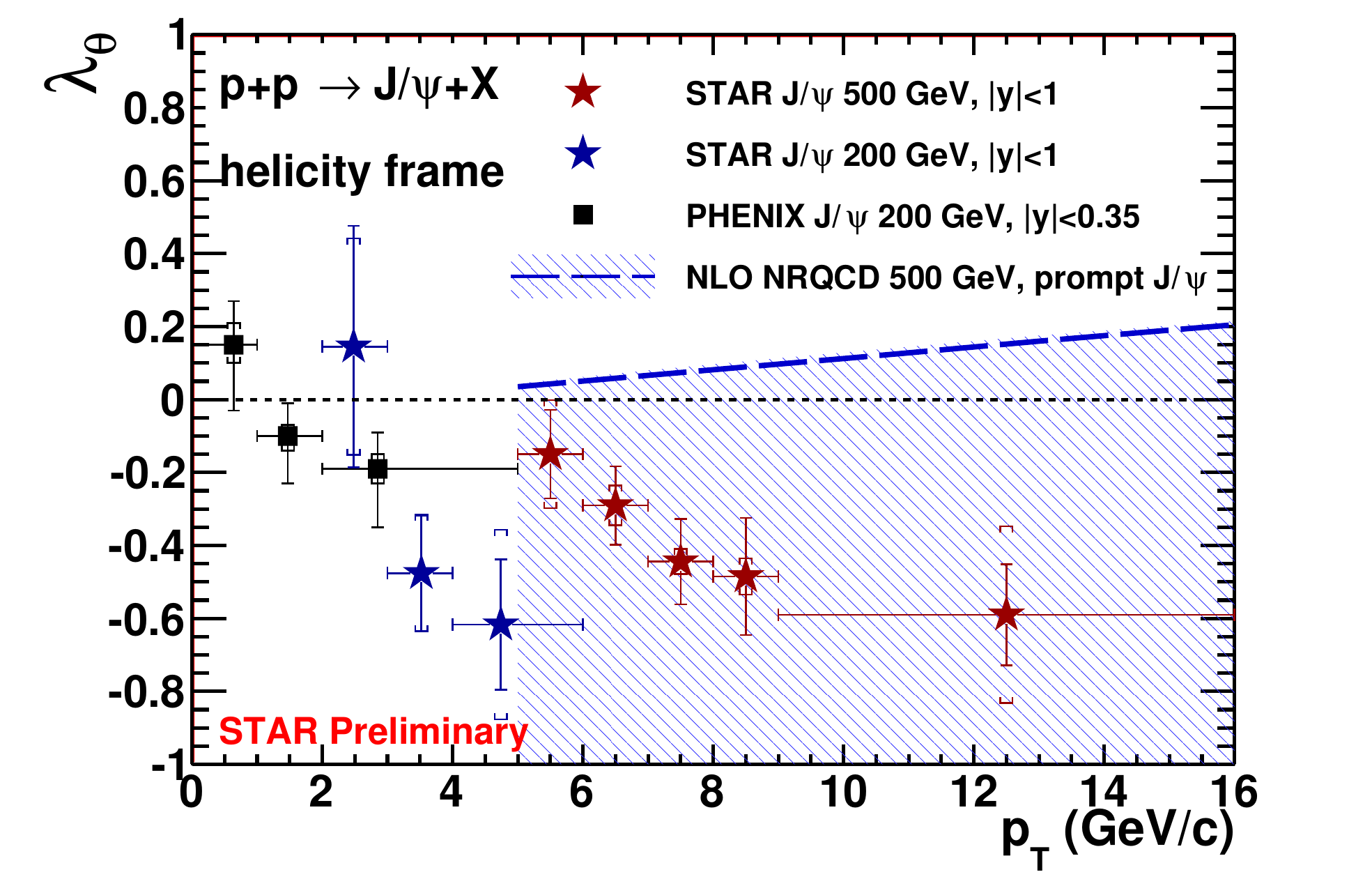}
		\caption{Left: J/$\psi$ polarization parameter $\lambda_{\theta}$ {\it{vs}} $p_{T}$ at $|y| <$ 1~\cite{Adamczyk:2013vjy} compared to the PHENIX measurement~\cite{Adare:2009js} and two model predictions~\cite{Chung:2009xr,Lansberg:2010vq}. Right: J/$\psi$ polarization parameter $\lambda_{\theta}$ {\it{vs}} $p_{T}$ at $\sqrt{s} =$ 500 GeV at mid-rapidity compared to results at $\sqrt{s} =$ 200 GeV and to the NLO NRQCD calculations~\cite{Chao:2012iv,Shao:2012fs,Shao:2014fca,Shao:2014yta}}
		\label{fig:lambdaTheta_HX}
\end{figure}

STAR has performed the first J/$\psi$ polarization measurement at $\sqrt{s} =$ 200 GeV at mid-rapidity and 2 $ < p_{T} <$ 6 GeV/$c$, in the HX frame~\cite{Adamczyk:2013vjy}. The left panel of Fig.~\ref{fig:lambdaTheta_HX} shows the $p_{T}$ dependence of $\lambda_{\theta}$. The STAR result is shown together with the PHENIX measurement at the same energy and in a lower $p_{T}$ range~\cite{Adare:2009js}. The results are also compared to LO NRQCD calculations (COM)~\cite{Chung:2009xr} and the NLO$^{+}$ CSM predictions~\cite{Lansberg:2010vq}. The trend observed in the RHIC data is towards longitudinal polarization as $p_{T}$ increases and, within experimental and theoretical uncertainties, the result is consistent with the NLO$^{+}$ CSM prediction.

\begin{figure}[ht]
		\centering
		\includegraphics[width=0.49\textwidth]{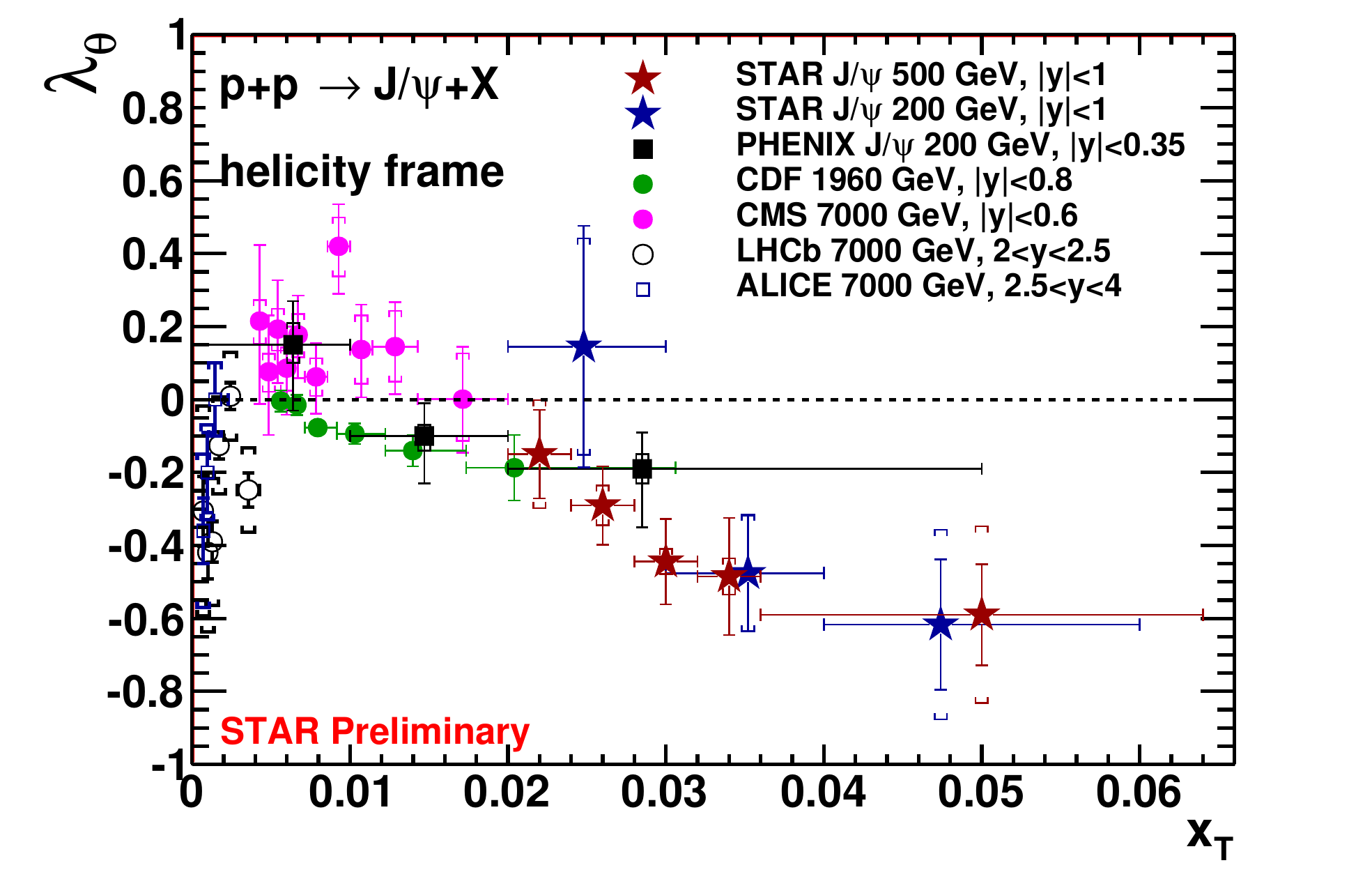}
		\includegraphics[width=0.49\linewidth]{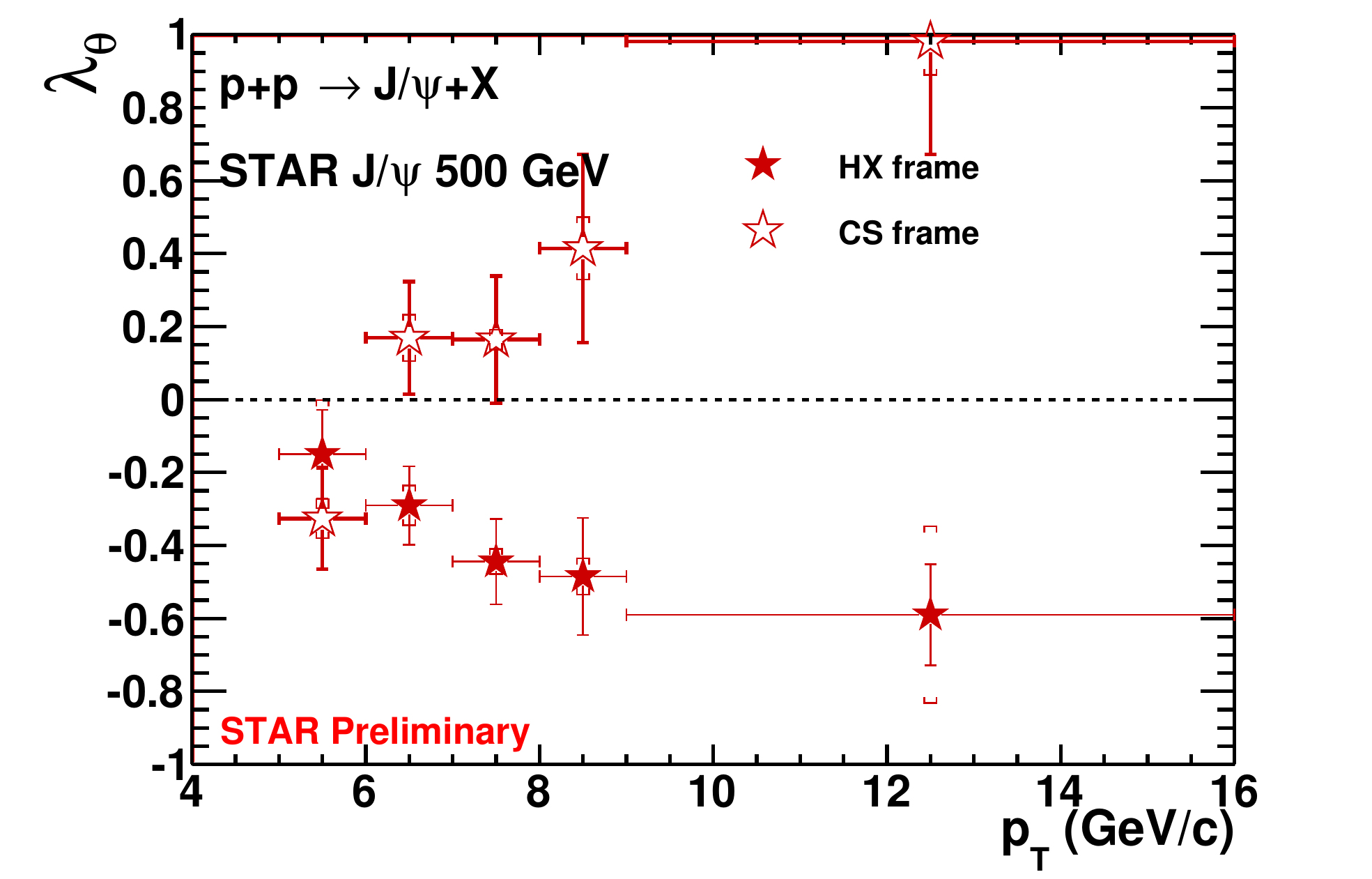}
		\caption{Left: $\lambda_{\theta}$ {\it{vs}} $x_{T} = 2p_{T}/\sqrt{s}$ in the HX frame. Right: $\lambda_{\theta}$ {\it{vs}} $p_{T}$ in the HX (closed symbols) and CS frame (open symbols).}
		\label{fig:lambdaTheta_HXxt}
\end{figure}

Due to limited statistics of the dataset at $\sqrt{s} =$ 200 GeV, only the $\lambda_{\theta}$ parameter related to the polar anisotropy has been extracted. 
The J/$\psi$ polarization measurements have been extended to a wider $p_{T}$ range of 5-16 GeV/$c$ using $p$+$p$ dataset at $\sqrt{s} =$ 500 GeV.
$\cos\theta$ and $\phi$ distributions integrated over $\phi$ and $\cos\theta$, respectively, are fitted simultaneously with Eq.~\ref{eq:decayD_cos},~\ref{eq:decayD_phi} in order to obtain $\lambda_{\theta}$ and $\lambda_{\phi}$ coefficients in each analyzed $p_{T}$ bin.
The same trend of $\lambda_{\theta}$ towards strong negative values is observed at $\sqrt{s} =$ 500 GeV as at $\sqrt{s} =$ 200 GeV in the HX frame. The results are shown in the right panel of Fig.~\ref{fig:lambdaTheta_HX} and are also compared to a NLO NRQCD prediction~\cite{Chao:2012iv,Shao:2012fs,Shao:2014fca,Shao:2014yta} at $\sqrt{s} =$ 500 GeV for the prompt J/$\psi$ production. The lower limit for $\lambda_{\theta}$ is unconstrained within the NRQCD calculations and the new data points should help to constrain color-octet Long-Distance Matrix Elements for the model.
Since similar trend of $\lambda_{\theta}$ as a function of $p_{T}$ is observed at RHIC for both analyzed colliding energies and two analyzed $p_{T}$ ranges, a comparison of the parameter values at different colliding energies as a function of $x_{T} = 2p_{T}/\sqrt{s}$ has been performed. The left panel of Fig.~\ref{fig:lambdaTheta_HXxt} shows $\lambda_{\theta}$ {\it{vs}} $x_{T}$ from RHIC, Tevatron and LHC experiments, and a common trend of $\lambda_{\theta}$ decreasing with increasing $x_{T}$ is observed.
 
\begin{figure}[ht]
		\centering
		\includegraphics[width=0.49\textwidth]{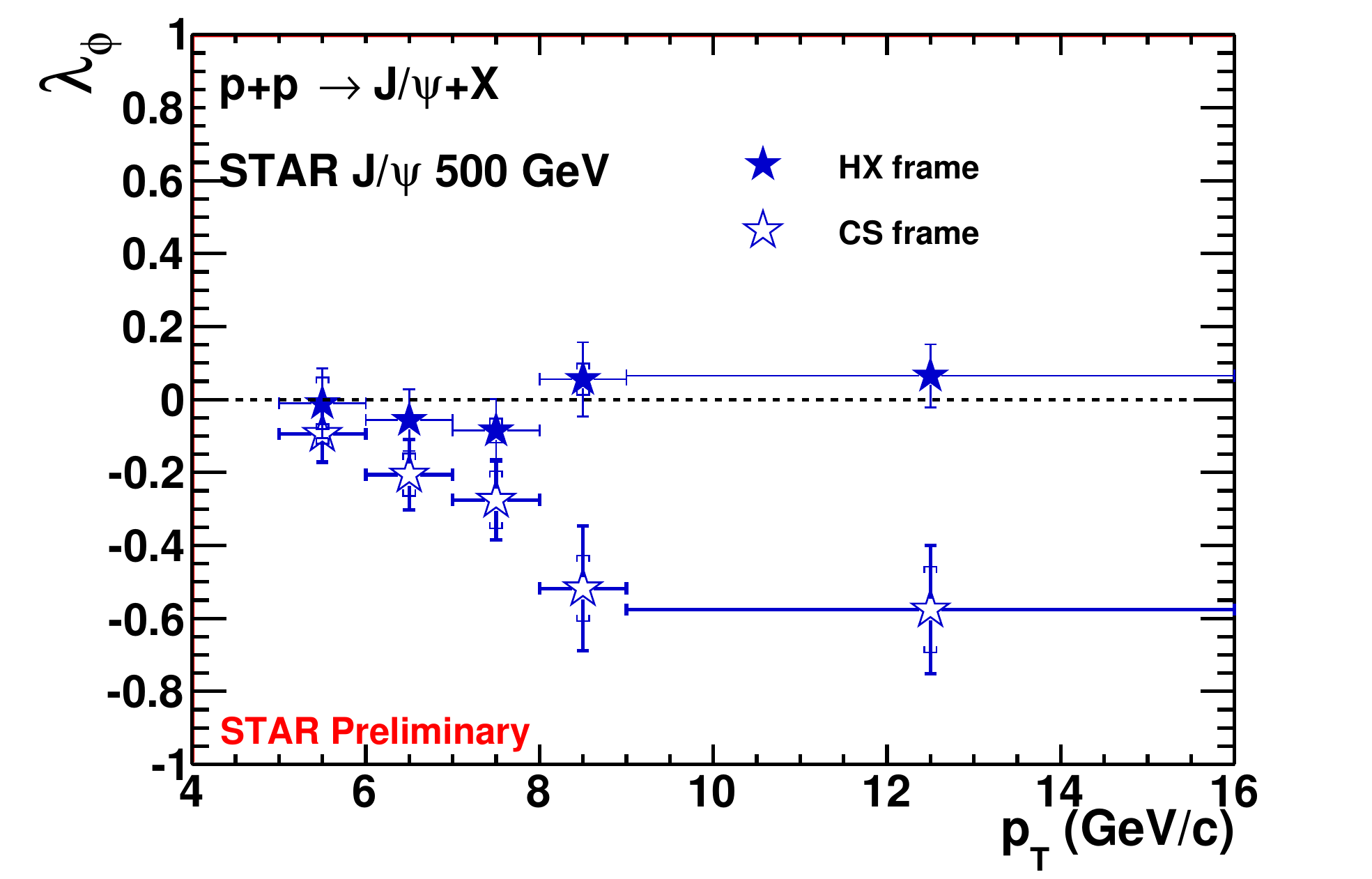}
		\includegraphics[width=0.49\linewidth]{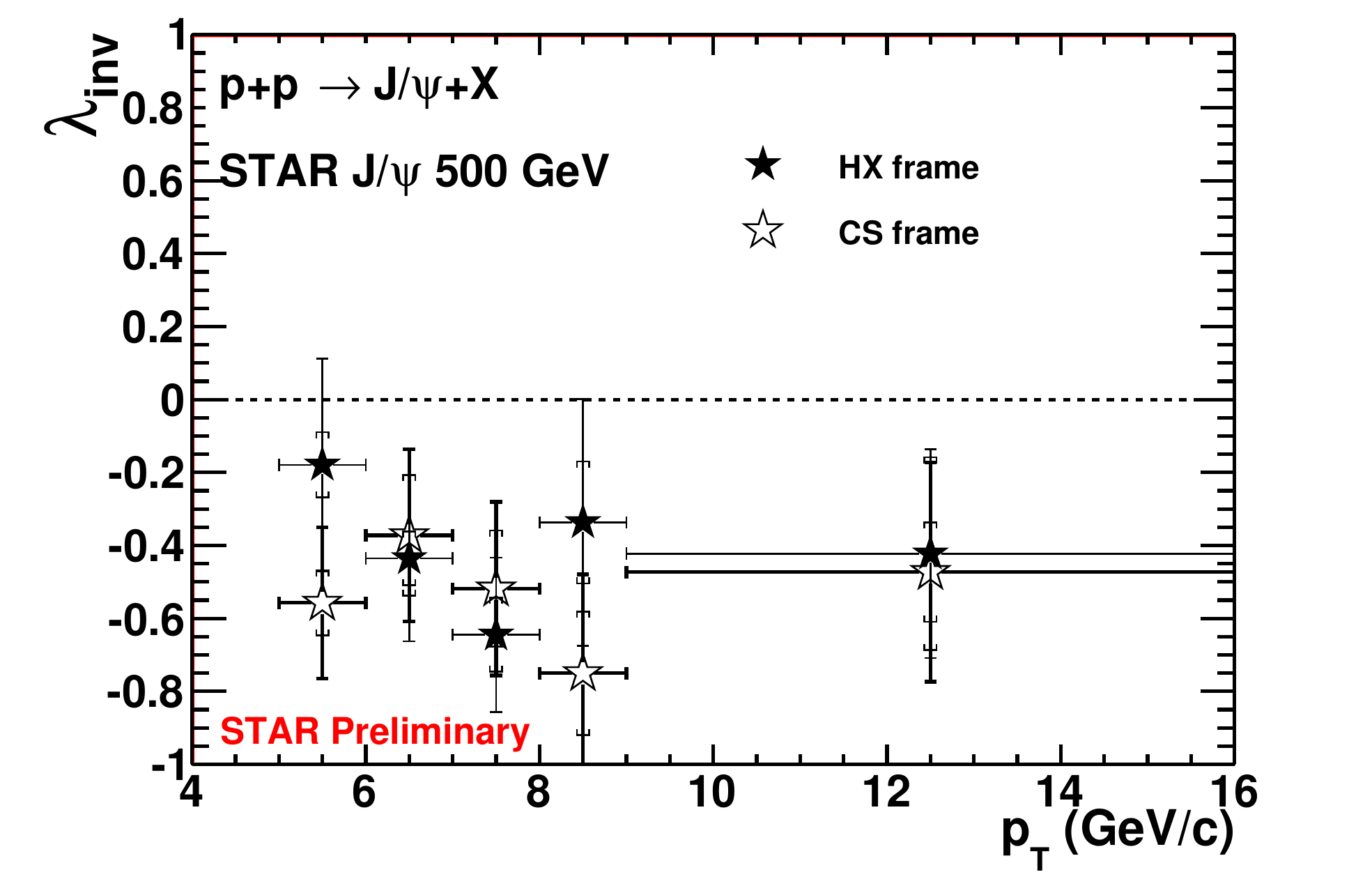}
		\caption{Left: $\lambda_{\phi}$ {\it{vs}} $p_{T}$ in the HX (closed symbols) and CS frame (open symbols). Right: $\lambda_{inv}$ {\it{vs}} $p_{T}$ in the HX (closed symbols) and CS frame (open symbols).}
		\label{fig:lambda_HXCS}
\end{figure}

Also, there is no strong azimuthal anisotropy observed in the HX frame at $\sqrt{s} =$ 500 GeV. Measured values of the $\lambda_{\phi}$ parameter are consistent with zero, as is shown in the left panel of Fig.~\ref{fig:lambda_HXCS}. To cross-check the measurement, both $\lambda_{\theta}$ and $\lambda_{\phi}$ parameters have been extracted in the CS frame in addition to the HX frame. The $\lambda_{\theta}$ is shown in the right panel of Fig.~\ref{fig:lambdaTheta_HXxt} and the $\lambda_{\phi}$ is depicted in the left panel of Fig.~\ref{fig:lambda_HXCS}. Results in the CS frame are shown as open symbols and compared to the results in the HX frame presented as solid points. We observe different $p_{T}$ dependence of the parameters. However, the frame invariant parameters, $\lambda_{inv}$, are in agreement between the frames and $\lambda_{inv}$ has negative values in both frames, as shown in the right panel of Fig.~\ref{fig:lambda_HXCS}.

\section{Summary}
\label{sec:summary}

STAR has measured J/$\psi$ polarization at $\sqrt{s}$ = 200 and 500 GeV at mid-rapidity.
At both energies a trend of $\lambda_{\theta}$ parameter towards negative values with increasing $p_{T}$ is observed in the helicity frame. The result at $\sqrt{s}$ = 200 GeV is consistent with the NLO$^{+}$ CSM prediction. Also, a common $x_{T}$ dependence of RHIC, Tevatron and LHC $\lambda_{\theta}$ measurements is observed.
At $\sqrt{s}$ = 500 GeV there is no azimuthal anisotropy seen in the helicity frame: $\lambda_{\phi}$ values are consistent with zero. The 500 GeV result has been also cross-checked by performing polarization analysis in the Collins-Soper frame. Different trends of $\lambda_{\theta}$ and $\lambda_{\phi}$ parameter values are observed but the frame invariant parameters $\lambda_{inv}$ are in agreement between the frames.

\section*{Acknowledgements}

This publication was supported by the European social fund within the framework of realizing the project ,,Support of inter-sectoral mobility and quality enhancement of research teams at Czech Technical University in Prague'', CZ.1.07/2.3.00/30.0034 and by Grant Agency of the Czech Republic, grant No.13-20841S.

\bibliographystyle{JHEP} 
\bibliography{EPSbib.bib}

\providecommand{\href}[2]{#2}\begingroup\raggedright\begin{thebibliography}{10}

\bibitem{Abelev:2009qaa}
{\bf STAR} Collaboration, B.~I. Abelev et~al., {\it {J/psi production at high
  transverse momentum in p+p and Cu+Cu collisions at s(NN)**1/2 = 200GeV}},
  {\em Phys. Rev.} {\bf C80} (2009) 041902,
  [\href{http://arxiv.org/abs/0904.0439}{{\tt arXiv:0904.0439}}].

\bibitem{Adamczyk:2012ey}
{\bf STAR} Collaboration, L.~Adamczyk et~al., {\it {$J/\psi$ production at high
  transverse momenta in $p+p$ and Au+Au collisions at $\sqrt{s_{NN}} = 200$
  GeV}},  {\em Phys. Lett.} {\bf B722} (2013) 55--62,
  [\href{http://arxiv.org/abs/1208.2736}{{\tt arXiv:1208.2736}}].

\bibitem{Adare:2011vq}
{\bf PHENIX} Collaboration, A.~Adare et~al., {\it {Ground and excited
  charmonium state production in $p+p$ collisions at $\sqrt{s}=200$ GeV}},
  {\em Phys. Rev.} {\bf D85} (2012) 092004,
  [\href{http://arxiv.org/abs/1105.1966}{{\tt arXiv:1105.1966}}].

\bibitem{PhysRevLett.79.572}
{\bf CDF} Collaboration, F.~Abe et~al., {\it {$J/\psi$ and $\psi(2S)$
  production in $p\bar{p}$ collisions at $\sqrt{s} = 1.8$ TeV}},  {\em Phys.
  Rev. Lett.} {\bf 79} (1997) 572--577.

\bibitem{Acosta:2004yw}
{\bf CDF} Collaboration, D.~Acosta et~al., {\it {Measurement of the $J/\psi$
  meson and $b-$hadron production cross sections in $p\bar{p}$ collisions at
  $\sqrt{s} = 1960$ GeV}},  {\em Phys. Rev.} {\bf D71} (2005) 032001,
  [\href{http://arxiv.org/abs/hep-ex/0412071}{{\tt hep-ex/0412071}}].

\bibitem{Aad:2011sp}
{\bf ATLAS} Collaboration, G.~Aad et~al., {\it {Measurement of the differential
  cross-sections of inclusive, prompt and non-prompt $J/\psi$ production in
  proton-proton collisions at $\sqrt{s}=7$ TeV}},  {\em Nucl. Phys.} {\bf B850}
  (2011) 387--444, [\href{http://arxiv.org/abs/1104.3038}{{\tt
  arXiv:1104.3038}}].

\bibitem{Khachatryan:2010yr}
{\bf CMS} Collaboration, V.~Khachatryan et~al., {\it {Prompt and non-prompt
  $J/\psi$ production in $pp$ collisions at $\sqrt{s}=7$ TeV}},  {\em Eur.
  Phys. J.} {\bf C71} (2011) 1575, [\href{http://arxiv.org/abs/1011.4193}{{\tt
  arXiv:1011.4193}}].

\bibitem{Aaij:2011jh}
{\bf LHCb} Collaboration, R.~Aaij et~al., {\it {Measurement of $J/\psi$
  production in $pp$ collisions at $\sqrt{s}=7~\rm{TeV}$}},  {\em Eur. Phys.
  J.} {\bf C71} (2011) 1645, [\href{http://arxiv.org/abs/1103.0423}{{\tt
  arXiv:1103.0423}}].

\bibitem{Ackermann:2002ad}
{\bf STAR} Collaboration, K.~H. Ackermann et~al., {\it {STAR detector
  overview}},  {\em Nucl. Instrum. Meth.} {\bf A499} (2003) 624--632.

\bibitem{Anderson:2003ur}
M.~Anderson et~al., {\it {The Star time projection chamber: A Unique tool for
  studying high multiplicity events at RHIC}},  {\em Nucl. Instrum. Meth.} {\bf
  A499} (2003) 659--678, [\href{http://arxiv.org/abs/nucl-ex/0301015}{{\tt
  nucl-ex/0301015}}].

\bibitem{Beddo:2002zx}
{\bf STAR} Collaboration, M.~Beddo et~al., {\it {The STAR barrel
  electromagnetic calorimeter}},  {\em Nucl. Instrum. Meth.} {\bf A499} (2003)
  725--739.

\bibitem{Llope:2012zz}
{\bf STAR} Collaboration, W.~J. Llope, {\it {Multigap RPCs in the STAR
  experiment at RHIC}},  {\em Nucl. Instrum. Meth.} {\bf A661} (2012)
  S110--S113.

\bibitem{Faccioli:2010kd}
P.~Faccioli, C.~Lourenco, J.~Seixas, and H.~K. Wohri, {\it {Towards the
  experimental clarification of quarkonium polarization}},  {\em Eur. Phys. J.}
  {\bf C69} (2010) 657--673, [\href{http://arxiv.org/abs/1006.2738}{{\tt
  arXiv:1006.2738}}].

\bibitem{Adamczyk:2013vjy}
{\bf STAR} Collaboration, L.~Adamczyk et~al., {\it {$J/\psi$ polarization in
  p+p collisions at $\sqrt{s}$ = 200 GeV in STAR}},  {\em Phys. Lett.} {\bf
  B739} (2014) 180--188, [\href{http://arxiv.org/abs/1311.1621}{{\tt
  arXiv:1311.1621}}].

\bibitem{Adare:2009js}
{\bf PHENIX} Collaboration, A.~Adare et~al., {\it {Transverse momentum
  dependence of J/psi polarization at midrapidity in p+p collisions at s**(1/2)
  = 200-GeV}},  {\em Phys. Rev.} {\bf D82} (2010) 012001,
  [\href{http://arxiv.org/abs/0912.2082}{{\tt arXiv:0912.2082}}].

\bibitem{Chung:2009xr}
H.~S. Chung, C.~Yu, S.~Kim, and J.~Lee, {\it {Polarization of prompt J/psi in
  proton-proton collisions at RHIC}},  {\em Phys. Rev.} {\bf D81} (2010)
  014020, [\href{http://arxiv.org/abs/0911.2113}{{\tt arXiv:0911.2113}}].

\bibitem{Lansberg:2010vq}
J.~P. Lansberg, {\it {QCD corrections to J/psi polarisation in pp collisions at
  RHIC}},  {\em Phys. Lett.} {\bf B695} (2011) 149--156,
  [\href{http://arxiv.org/abs/1003.4319}{{\tt arXiv:1003.4319}}].

\bibitem{Chao:2012iv}
K.-T. Chao, Y.-Q. Ma, H.-S. Shao, K.~Wang, and Y.-J. Zhang, {\it {$J/\psi$
  Polarization at Hadron Colliders in Nonrelativistic QCD}},  {\em Phys. Rev.
  Lett.} {\bf 108} (2012) 242004, [\href{http://arxiv.org/abs/1201.2675}{{\tt
  arXiv:1201.2675}}].

\bibitem{Shao:2012fs}
H.-S. Shao and K.-T. Chao, {\it {Spin correlations in polarizations of P-wave
  charmonia $\chi_{cJ}$ and impact on $J/\psi$ polarization}},  {\em Phys.
  Rev.} {\bf D90} (2014), no.~1 014002,
  [\href{http://arxiv.org/abs/1209.4610}{{\tt arXiv:1209.4610}}].

\bibitem{Shao:2014fca}
H.-S. Shao, Y.-Q. Ma, K.~Wang, and K.-T. Chao, {\it {Polarizations of
  $\chi_{c1}$ and $\chi_{c2}$ in prompt production at the LHC}},  {\em Phys.
  Rev. Lett.} {\bf 112} (2014), no.~18 182003,
  [\href{http://arxiv.org/abs/1402.2913}{{\tt arXiv:1402.2913}}].

\bibitem{Shao:2014yta}
H.-S. Shao, H.~Han, Y.-Q. Ma, C.~Meng, Y.-J. Zhang, and K.-T. Chao, {\it
  {Yields and polarizations of prompt $J/\psi$ and $\psi(2S)$ production in
  hadronic collisions}},  {\em JHEP} {\bf 05} (2015) 103,
  [\href{http://arxiv.org/abs/1411.3300}{{\tt arXiv:1411.3300}}].

\end{thebibliography}\endgroup


\end{document}